\documentclass[apj]{emulateapj}
\usepackage{apjfonts}
\submitted{Accepted by ApJ July 9, 2003}
 
\newcommand{\egret}{{\em EGRET}\/}

\newcommand{\likeprog}{{\tt LIKE}\/}

\newcommand{\etal}{et~al.}

\newcommand{\eq}[1]{equation~(\ref{#1})}

\newcommand{\fig}[1]{Figure~\ref{#1}}
\newcommand{\sect}[1]{\S\ref{#1}}

\def\aisr{Adv. Space Res.}
\def\aj{AJ}			
		
\def\apj{ApJ}			
\def\apjl{ApJ}		
\def\apjs{ApJS}		
\def\ao{Appl.Optics}		
		
\def\aap{A\&A}		
		
\def\aaps{A\&AS}

\def\mnras{MNRAS}

\def\prd{Phys.Rev.D}		
\def\prl{Phys.Rev.Lett}

\def\deg{\hbox{$^\circ$}}

\def\earth{\hbox{\footnotesize $\oplus$}}

\begin{document}
\bibliographystyle{apj}

\title{Variability of EGRET Gamma-Ray Sources}
\shorttitle{Variability of EGRET Gamma-Ray Sources}
\author{P. L. Nolan}
\affil{W. W. Hansen Laboratory, Stanford University, Stanford CA 94305-4085, 
USA}
\email{Patrick.Nolan@stanford.edu}
\author{W. F. Tompkins}
\affil{Geophysical Associates, 324 Franklin St., Harrisonburg VA 22801, USA}
\email{wtompkins@icarion.com}
\author{I. A. Grenier}
\affil{Service d'Astrophysique, Centre d'\'Etudes de Saclay, 91991 
Gif/Yvette,  France}
\email{isabelle.grenier@cea.fr}
\author{P. F. Michelson}
\affil{Physics Department and W. W. Hansen Laboratory,\\ Stanford University, 
Stanford CA 94305-4060, USA}
\email{peterm@leland.stanford.edu}

\begin{abstract}
The variability of the high-energy gamma ray sources in the Third
\egret\ catalog is analyzed by a new method.  We re-analyze the
\egret\ data to calculate a likelihood
function for the flux of each source in each observation, both
for detections and upper limits.  These 
functions can be combined in a uniform manner with a simple model of
the flux distribution to characterize the flux variation by a
confidence interval for the
relative standard deviation of the flux.  The main
result is a table of these values for almost all the cataloged
sources.  As expected, the identified pulsars are steady emitters and
the blazars are mostly highly variable.  The unidentified sources
are heterogeneous, with greater variation at higher Galactic latitude.
There is an indication that pulsar wind nebulae are associated with
variable sources.
There is a population of variable sources along the Galactic plane,
concentrated in the inner spiral arms.
\end{abstract}


\section{Introduction}

Of the 271 point sources of high-energy gamma-rays detected by
the Energetic Gamma Ray Experiment Telescope (\egret) on the Compton Gamma
Ray Observatory spacecraft,
most remain unidentified~\citep{cat3}.  The nature of these sources is
one of the biggest questions raised by the \egret\ mission.  
Certainly some of them will turn out
to belong to the established classes of sources: Active Galactic
Nuclei (AGN)~\citep{tomthesis,dse} 
and pulsars~\citep{romani,mukherjee_unid,merck_unid}.
A few, as well, may be discovered to be artifacts of the Galactic diffuse
emission.
However there are more unidentified sources than expected from these
populations~\citep{merck_unid}.  
The large error regions for the \egret\ sources often 
make it difficult to discern which of several candidates is the true
counterpart in another wavelength.

When studying the unidentified sources, there are several observables
which one might use to determine their nature.  One can look at the
spatial 
distribution~\citep{kanbach_characteristics,grenier} 
to estimate what fraction
of the population is Galactic in nature, and what scale height the Galactic
fraction has.  One can look at positional correlations with other object
types and Galactic structures~\citep{cat2,cat3,romero,gehrels,grenier2001,
sturner_snr,esposito_snr,distantSNR,Mattox97,Wallace02,gasClouds,
ClusterCoinc,ClusterUpper}.  
The energy spectrum is
another characteristic which can be used to classify sources as
``pulsar-like'', ``AGN-like'', or ``other''~\citep{merck_unid}.

In this same vein, one can examine the \egret\ catalog sources for evidence
of time variability, in hopes of distinguishing the various source classes.
The known pulsars are seen to have a fairly constant flux ($\pm<10\%$ 
when averaged over many pulse periods, consistent with the
systematic uncertainties in \egret\ flux measurements), 
as is expected from the nature of their energy production.
Many AGN are seen to flare dramatically, which makes their identification
easier.
But, as with many aspects of the \egret\ data, it is difficult to
characterize the time variability of most sources, because of the limited statistics
and the non-continuous observations.  Thus it is natural
to look for a statistic which will measure the variability in a rigorous way.

This work is based on the dissertation of \citet{billthesis}.
Changes in the details of the method have caused changes in the
numerical results, but the conclusions are not altered.

In \sect{PREV_METH} we briefly review the previous methods which have been
used for evaluating the variability of \egret\ point sources.
In \sect{OURWAY} we outline an improved method.
In \sect{ANAL_SECTION} we describe how this method was applied to the
\egret\ data.
The results are described in \sect{VARIRESULTS} and summarized briefly in
\sect{CONCLUSIONS}.

\section{Previous Methods\label{PREV_METH}}

All previous analyses of the variability of \egret\ sources have used the flux
values published in the Second \citep{cat2} or Third \egret\ 
Catalog \citep{cat3}, 
hereafter called 2EG and 3EG.  This introduces the problem of upper limits.  
Many of the individual observations produced no positive detections.  
For these observations, the flux is reported as a 95\% confidence upper 
limit produced
by \likeprog, the standard \egret\ likelihood program \citep{like}.

The first comprehensive method used for estimating the variability of 
\egret\ sources is the one introduced by \citet{mclaughlin}.  This method 
finds the $\chi^2$ of the measured source intensities in 2EG, assuming
a constant flux, and uses
$V = - \log Q$, where $Q$ is the tail probability of obtaining such a large
$\chi^2$.
There are two problems with this method: one statistical, and one
conceptual.  The statistical problem is introduced when using upper
limits:  \citet{mclaughlin} use zero as the flux estimate when the 
measured flux estimate would be negative, and for the error they use the 
\likeprog\ upper limits.
Thus their resulting ``$\chi^2$'' will not have a $\chi^2$ distribution.
Since the \likeprog\ results are statistically conservative, this means that
sources with upper limits included in the analysis will have a lower
$V$ than they should.
The conceptual problem with the $V$ method is the use of a $\chi^2$ statistic
to determine variability.  The $Q$ statistic is the p-value for
the hypothesis of constant source flux.  Thus, sources
with a large $V$ are inconsistent with being constant.  But we do not
know whether such a source has a large $V$ because of large intensity
fluctuations, or because of small error bars on the intensity measurements.
Similarly, a source with a small $V$ might truly be constant, or it might
just have very poor measurements of its flux.

\citet{wallace} also applied this method to two-day segments of
\egret\ data, looking for rapid variation that would not have
been discovered by examining full viewing periods.  They were
severly limited by large statistical uncertainties, but they found
previously-undiscovered variation in several sources.

\citet{torres_num,torres} describe a statistic called the $I$ index.
$I$ is a normalized ratio of the standard deviation to the average flux.
This is somewhat similar to the approach we describe below.  It is a step 
toward dealing with the conceptual problem described above, but it 
introduces another ad-hoc approach to dealing with upper limits.
Also there is no estimate of confidence intervals for $I$ values.

Other, less quantitative criteria have been used to produce lists of sources
which are not highly variable.  These lists are heterogeneous collections
of sources with rather different properties.  In addition to truly
non-variable sources, they tend to include bright variable sources.
They exclude mainly the sources which flared once or twice and were
otherwise undetectable. For instance, \citet{grenier2001} defined a 
class of 88 {\em persistent} unidentified sources.
These are the ones which are significantly detected ($>4 \sigma$ or
$>5 \sigma$, depending on position) in the cumulative data up to 1995 October.
\citet{gehrels} produced a similar list of 120 {\em steady}, unidentified
 sources.  They required that the flux
in the cumulative data must be either a more significant detection than any
individual observation or within $3\sigma$ of the largest 
individual flux.  Again the effects of average brightness and variability
are hard to disentangle.  It is difficult to compare sources or
populations on the basis of these lists because they provide only
a binary variable/nonvariable classification.

\section{A Likelihood Approach\label{OURWAY}}
\subsection {What Do We Want to Measure?}

We propose to answer two questions about the variability of a source:
How much does the flux vary and how precisely do we know this?
To do this we define
\begin{equation}
\delta \equiv \frac{\sigma}{\mu},
\end{equation}
where $\mu$ and $\sigma$ are the average and standard deviation,
respectively, of the true flux of the source.  
This fractional variability is much closer to
what we really want to know about a source.  There could also be other 
statistics which
might be more meaningful for flaring sources, such as the peak flare flux
divided by the quiescent level.
Since the flux measurements have uncertainties, we must find a way 
to estimate $\delta$ from the data.
Unfortunately this can't be done in a completely model-independent way.

\subsection{Modeling the Source Flux Distribution}
We ignore the ordering and spacing of the individual flux
measurements.  We assume only that all flux values of a particular
source are random values drawn from the same probability distribution.
This approach elminates all consideration of long timescale correlations
in the data.
The probability distribution is parameterized by $\delta$.
Such a model is necessary in order to disentangle the effects of real
flux variation from statistical fluctuations.
An infinite number of source flux distributions exist which will 
yield the same $\delta$; we want to pick one that
is compatible with our notions of what the true source flux distribution
should be, but which is also fairly general.

The first source flux distribution one might try is a Gaussian.
Thus, we would find the likelihood of obtaining our data
given that the true flux $S$ was drawn from a Gaussian with mean $\mu$ and
standard deviation $\sigma = \mu\delta$.  We could then find
the maximum likelihood value of $\delta$, together with a confidence interval
for $\delta$ defined in the standard way \citep{eadie}.  Alternatively,
we could follow the Bayesian procedure: form priors for the distributions
of $\mu$ and $\delta$, and find an estimate of $\delta$, together with an
error region, by marginalizing over $\mu$.
The use of the Gaussian distribution for source fluxes has a flaw, however.
It allows the possibility of $\mu = 0$ ($\Rightarrow \delta = \infty$), as
well as the unphysical possibility of negative $\mu$.  

Instead of a Gaussian, we choose to use the
gamma distribution; that is, the flux $S$ is drawn from the differential 
probability distribution
\begin{equation}
P(S) = \frac{S^{\alpha-1}e^{-S/\beta}}{\beta^\alpha \Gamma(\alpha)},
\end{equation}
where the adjustable parameters $\alpha\,(>0)$ and $\beta\,(>0)$ describe
the shape and scaling of the function, respectively.
The parameters can be related to the desired quantities by
$\mu = \beta\alpha$, $\sigma=\beta\sqrt{\alpha}$, and
$\delta=\alpha^{-1/2}$.
This distribution function has a peak if $\alpha>1$. For $0<\alpha<1$, it decreases
monotonically from $S=0$.  This behavior is a good match for the different
types of observed variations.  Some sources are steady (large $\alpha$), 
while many others
are mostly faint with occasional flares (small $\alpha$).

\citet{billthesis} did the same analysis described below using a 
lognormal distribution.  There is an excellent correlation between the
results, but the gamma distribution produced fewer numerical difficulties,
such as lack of convergence.  We take this as a validation of our
choice of probability distribution.

\subsection{Characterizing the Single-Measurement Likelihood \label{intcurve}}

It is standard in \egret\ analysis to calculate the likelihood of the observed
data for a certain source flux \citep{like}.  Typically, the
values output by an analysis program are the most likely flux and a
confidence interval.  In the case of upper limits, the most likely flux is
taken as zero, and the confidence interval might be defined in one of several
ways \citep{pdg98}.

For this work, we would like to be able to calculate the likelihood for
any source flux, not necessarily one near the maximum likelihood
value.  Thus it is useful to look for a parameterized family of curves
which will closely fit the full likelihood function.  
The results of the many individual measurements can be stored in a compact
form and reproduced quickly as needed.

After examining many such curves, we found a simple function which fulfills our
requirements.  We choose
\begin{equation}
{\cal L}(S) = e^{-(k+S/S_*)} (k+S/S_*)^p,
\label{intcurvefit}
\end{equation}
where $S_*$, $k$, and $p$ are adjustable parameters and only $S\geq 0$ is allowed.
A likelihood function need not be normalized.
If $p>k$, ${\cal L}(S)$ has a peak at $S=(p-k)S_*$ of width
$S_*\sqrt{p}$, and the \egret\ test statistic \citep{like} $TS = 
2p(k/p -\ln(k/p)-1)$.
If $p\leq k$, ${\cal L}(S)$ has its maximum at $S=0$.
This function was originally derived by analogy with the Poisson
probability function.  The parameter $p$ plays the role of an effective
number of detected photons, while $1/S_*$ is the exposure and the offset
$k$ accounts for the effects of diffuse emission and nearby point sources.
$\mathcal{L}(S)$ would become singular if $k < 0$, but this never occurred in 
our analysis.

\begin{figure}
\centering
\plotone{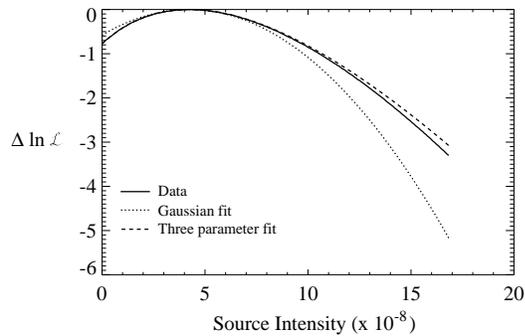}
\caption[Log Likelihood vs Intensity]{\label{intcurvefig}
The likelihood curve for a sample source, together with a Gaussian of the
same width, and a curve of the form in \eq{intcurvefit}.
}
\end{figure}

This form fits the observed likelihood functions rather well
(\fig{intcurvefig}), both in cases of large source flux and in upper 
limit situations.

\subsection{Calculating Likelihood \label{taucalc}}

With the parameterized distribution of source fluxes from the preceding
section, we can use Bayes's Theorem to write the likelihood of observing a given 
viewing period's
worth of data ($D_i$) from a source with a certain $\mu$ and $\delta$:
\begin{eqnarray}
{\cal L}(\mu, \delta, D_i) & = & \int_0^\infty dS_i\, P(D_i | S_i) P(S_i|\mu, \delta) \\
& = &  \int_0^\infty dS_i\, \mathcal{L}(S_i)\,P(S_i|\mu,\delta) \\
& = &  \int_0^\infty dS_i\, e^{-(k_i+S_i/S_{*i})}(k_i+S_i/S_{*i})^{p_i} \nonumber \\
 & & \times \frac{1}{(\mu\delta^2)^{1/\delta^2}\Gamma(1/\delta^2)}
     S_i^{(1/\delta^2-1)} e^{-S_i/(\mu\delta^2)},
\label{tauonevp}
\end{eqnarray}
where the second term in \eq{tauonevp} is the gamma distribution
written in terms of $\mu$ and $\delta$.

The likelihood of a sequence of observations ($D_1 \dots D_M$) will then
be 
\[
{\cal L}(\mu, \delta) = \prod_{i=1}^M {\cal L}(\mu, \delta, D_i),
\]
or
\begin{equation}
\ln {\cal L}(\mu, \delta) = \sum_{i=1}^M \ln {\cal L}(\mu, \delta, D_i).
\end{equation}

\subsection{Dealing with Time Scales \label{timescales}}

We have assumed that the source flux during all the measurements
was drawn independently from the same distribution.  This can be problematic
if the measurements last for varying lengths of time.  For example, in the
\egret\ data, there are frequently several successive viewing periods of
a few days duration, pointing in nearly the same direction.  This is in
contrast to the normal mode of pointing in one direction for two weeks at
a time.  We do not want our results to depend on whether or not a two week
viewing period was chopped into several bits.

Our solution is to combine successive measurements
if they occur within 30 days, and treat them as one measurement.
Instead of the single-observation likelihood in \eq{intcurvefit},
we use a composite $\mathcal{L}(S)$ which is the product of the
individual terms.  The generalization of \eq{tauonevp} is obvious.
These combined measurements are reflected in the number of observations 
reported in Table~1.

Thus this analysis is sensitive to variations with a time scale of a month
or more.  We will not be able to detect the 1-2 day flares found by
\citet{wallace} from several \egret\ sources.

\section{Analysis\label{ANAL_SECTION}}

\subsection{Method \label{method}}

The first step in this analysis was the re-analysis of the 3EG
catalog with a modified likelihood program which produces a likelihood
function in the form of \eq{intcurvefit} \citep{billthesis}.  
For each Viewing Period, any of the 271 sources from 3EG Catalog which 
were within the instrument's field
of view were included in the source list.  In addition, any of the 145 marginal
sources (which were used in the original analysis of the catalog but
were not significant enough to merit inclusion in the official list) which
were in the field of view were added as well.  The maximum likelihood
set of source fluxes was then found.  For each source, the parameters
characterizing ${\cal L}(S)$ described in \sect{intcurve} were found.
All source measurements were of flux ($> 100\ {\rm MeV}$), assuming a spectral
index of $-2.0$, using all photons within $15 \deg$ of each source, and 
including all nearby sources (both catalog and marginal) in the model.  
We use only data taken with the full \egret\ field of view, which includes
everything through Viewing Period 429.0 (ending 1995 September 27)
except for four periods (403.0, 403.5, 411.1, and 411.5).
With the smaller field of view used to take the rest of the data,
we are not confident that the detector response functions are
calibrated well enough for our purpose.  Also the number of point
sources detected in each observation is reduced, so the remaining 
data are less useful.
Results are used only for the point sources within $25\deg$ of the
center of the field of view.  This data selection is nearly the
same as that used to produce the 3EG Catalog.

The likelihood of the sequence of observations as a function of $\delta$
was then calculated, as in \sect{taucalc}.  All observations within one
month of each other were lumped together, as per \sect{timescales}.
From this likelihood, the most likely values of $\delta$ were obtained,
as well as confidence intervals.  The mean flux $\mu$ can be marginalized
away as a nuisance parameter after checking that its most likely value
is consistent with the 3EG flux.

\subsection{Sources Omitted from This Analysis\label{omissions}}

In the results reported below, several 3EG sources are omitted for various
reasons.

Seven sources near the bright \objectname{Crab} and \objectname{Vela} 
pulsars are omitted because
they are believed to be artifacts \citep{artifacts}:  
\objectname{3EG J0824$-$4610},
\objectname{3EG J0827$-$4247}, \objectname{3EG J0828$-$4954}, 
\objectname{3EG J0841$-$4356}, \objectname{3EG J0848$-$4429},
\objectname{3EG J0859$-$4247}, and \objectname{3EG J0521+2147}.

The solar flare \objectname{3EG J0516+2320} is omitted because it is clearly 
a unique occurrence.

The sources \objectname{3EG J0829+2413} and \objectname{3EG J1424+3734} 
each had only one observation
within 25\arcdeg\ of the center of the field of view.  Thus the variability 
of these sources can't be evaluated.

A cross-check of this analysis and the 3EG flux values turned up three sources
with serious discrepancies: \objectname{3EG J0556+0409},
\objectname{3EG J0628+1847}, and \objectname{3EG J1310$-$0517}.  
These three sources are omitted. 
The latter two are near very bright sources (\objectname{Geminga} and
\objectname{3C279}), so we believe that our modified 
likelihood program may become confused in such areas. 
It is encouraging that only these three sources fail this test.  
The consistency of results for all the others encourages us that
the modified likelihood program is valid in the other cases.

\subsection{Systematic Errors}

This analysis has two main sources of systematic error.  
The first is from the instrument itself.  
The sensitivity of the \egret\ instrument is 
variable \citep{cat2,cat3}.
Despite efforts to measure and compensate for long term drifts, 
there is a residual variability of about 10--11\% \citep{inflightcalibrate}.
In an effort to minimize this systematic error, only source observations
within 25\arcdeg\ of the instrument axis were used in the analysis.

It is conceivable that the variation in measured fluxes might be reduced 
by using the ``noisy'' measured sensitivity rather than the smoothed values
that are usually employed in \egret\ analysis.  We tested this
by calculating the fluxes of the three bright pulsars (Crab, Geminga,
Vela) by both methods.  With the measured sensitivity values, Vela
appeared much more variable, the Crab slighly more, and Geminga
slightly less.  Thus it appears to be appropriate to continue using
the smoothed sensitivities.

Because of the systemitic variability, the Galactic and isotropic background 
levels $G_M$ and $G_B$
were allowed to vary from viewing period to viewing period for a given
source in the likelihood analysis.  Otherwise, a small change in a 
strong background could lead to a large change in a source flux.
This procedure is standard practice for \egret\ data analysis \citep{like}.

The background levels were also allowed to float when ${\cal L}(S)$
was calculated.  Unfortunately, computational limitations precluded allowing
nearby sources to vary during this calculation as well.  Thus, nearby
sources were fixed at their maximum likelihood intensities (calculated
for that viewing period).  This will lead to an underestimation of the
error in flux, particularly for sources which have another source very close
by.

As both systematic effects will tend to exaggerate source variability,
it is expected that even constant sources will show some small variability.
To the extent that the systematics are dominated by the instrumental
effects, the measured variability will be equal to the true variability 
added in quadrature to a constant systematic variability of about 10\%.

\subsection{A Simple Variability Index\label{V12}}

Although the most likely value of $\delta$ and its confidence interval
provide a good description of a source's variability,
there is considerable interest in the simpler question of whether a source is variable
at all, independent of the question ``how much?''
This is the goal of \citet{mclaughlin}.
This problem is complicated by the presence of the systematic errors.
To help answer the question, we define the statistic $V_{12}$.  It
represents the confidence with which we can reject the hypothesis 
$\delta < 0.12$, where 0.12 is a conservative estimate of the average
systematic error.
\begin{equation}
V_{12} \equiv -\log_{10}(1 - P_{\chi^2}(r|1)),
\end{equation}
where
\begin{equation}
r = 2 \ln \frac{{\cal L}_{\rm max}(\mu,\delta)}{{\cal L}_{\rm max}(\mu,\delta=0.12)},
\end{equation}
${\cal L}_{\rm max}(\mu,\delta)$ is the unconstrained maximum likelihood,
${\cal L}_{\rm max}(\mu,\delta=0.12)$ is the maximum likelihood if $\delta$ is
constrained to equal the systematic value 0.12, and
$P_{\chi^2}(r|1)$ is the cumulative $\chi^2$ probability distribution with
one degree of freedom.  
The use of the $\chi^2$ distribution can be justified by Wilks's Theorem
\citep{cash} when several independent likelihood functions are combined,
even though they may not be Gaussian functions individually.
If the most likely $\delta$ is less than 0.12, then
there is no estimate for $V_{12}$.  
$V_{12}$ values greater than 16  are shown as $\infty$ in Table~1.
The p-value for rejecting the hypothesis that a source is 
non-variable is $10^{-V_{12}}$.

\section{Results \label{VARIRESULTS}}

The results of this analysis are shown in Table~1.
The columns in the table give the names in the 3EG catalog,
the most likely Galactic longitudes and latitudes, the most
likely values of $\delta$, 
the 68\% lower and upper confidence limits on $\delta$,
$V_{12}$, the number
of observations used in the analysis, possible associations with
source classes, and specific identifications.

\subsection{Variability by Source Class \label{classvar}}
As can be seen from Table~1, the limits on the variability of an 
individual
source are usually not very strict.  By grouping similar sources together,
one can learn more about the average behavior of a class of sources.
There are many sources which are found in multiple classes.
Many of the unidentified sources are found in densely-populated
regions containing various types of young-population objects.
The error boxes of the \egret\ sources often contain several
plausible counterparts.
Table~2 shows the results for the various source classes.
It contains a description of each class, the number of sources in
the class, the mean of $\delta$, and an estimate of the intrinsic RMS 
dispersion
of $\delta$ in excess of the statistical errors.
The last two quantities are calculated by assuming a Gaussian distribution
for $\delta$ and finding the likelihood as a function of mean and standard
deviation by a method similar to that used in \sect{method}.
\setcounter{table}{1}
\begin{deluxetable*}{lcccl}
\tablecaption{Variability of selected classes of sources}
\tablewidth{0pt}
\tablehead{
  \colhead{Source Class} & \colhead{Members} & \colhead{$\langle\delta\rangle$}
  & \colhead{RMS($\delta$)} & \colhead{Symbol}
}
\startdata
3EG Classification & & & & \\
\quad Pulsars & \phn\phn6 & $0.11 \pm 0.02$ & $< 0.07$ & PSR \\
\quad SNR associations & \phn10 & $0.27 \pm 0.09$ & $< 0.19$ & SNR \\
\quad AGN (``A'' in 3EG) & \phn67 & $0.70 \pm 0.08$ & $0.27 \pm 0.05$ & \\
\quad\quad Quasars & \phn51 & $0.77 \pm 0.09$ & $0.25 \pm 0.08$ & QSO\\
\quad\quad BL Lac objects & \phn15 & $0.49 \pm 0.16$ & $< 0.32$ & BLL\\
\quad Unidentified, $|b|<5\arcdeg$ & \phn47 & $0.42 \pm 0.06$ & $0.17 \pm 0.08$ & U0\\
\quad Unidentified, $5\arcdeg<|b|<15\arcdeg$ & \phn37 & $0.56 \pm 0.12$ & $0.23 \pm 0.09$ & U5\\
\quad Unidentified, $15\arcdeg<|b|<30\arcdeg$ & \phn45 & $0.42 \pm 0.12$ & $< 0.16$ & U15\\
\quad Unidentified, $|b| > 30\arcdeg$ & \phn29 & $0.84 \pm 0.15$ & $< 0.27$ & U30\\
Possible isolated neutron stars & & & & \\
\quad Geminga-like objects & \phn\phn3 & $0.13 \pm 0.11$ & $< 0.16$ & Gem \\
\quad Pulsars without PWN & \phn\phn7 & $0.21 \pm 0.15$ & $< 0.15$ & noPWN \\
\quad Pulsars with PWN & \phn\phn6 & $0.66 \pm 0.29$ & $0.28 \pm 0.16$ & PWN \\
Gould's Belt, unidentified, $|b|>5\arcdeg$ & \phn33 & $0.49 \pm 0.13$ & $< 0.26$ & G \\
Persistent, unidentified \citep{gehrels} & \phn88 & $0.32 \pm 0.05$ & $0.10 \pm 0.05$ & Per\\
Steady, unidentified \citep{grenier2001} & 120 & $0.35 \pm 0.05$ & $0.11 \pm 0.04$ & Std\\
Associations from \citet{romero} & & & & \\
\quad SNR & \phn17 & $0.28 \pm 0.10$ & $< 0.17$ & RSN \\
\quad WR stars & \phn\phn6 & $0.45 \pm 0.22$ & $< 0.27$ & WR \\
\quad Of stars & \phn\phn4 & $0.26 \pm 0.15$ & $< 0.19$ & Of\\
\quad OB associations & \phn22 & $0.40 \pm 0.10$ & $0.17 \pm 0.08$ & OB\\
\enddata
\end{deluxetable*}

To begin, consider four principal source classes based on the identifications
in the 3EG catalog:
Unidentified sources, Pulsars, Active Galactic Nuclei (AGN) (labeled
``A'' in 3EG), and sources which
are spatially coincident with Supernova Remnants (SNR).  The Unidentified source
class was divided according to the Galactic latitude.
These source classes were determined from the ``source ID'' (and by the
``Other Name'' category for the SNR associations) in the catalog.
Also, \objectname{3EG J1048$-$5840} (\objectname{PSR B1046$-$58}) has been  
added as a sixth member
of the pulsar class based on its recent detection \citep{kaspi2000}.  
Although pulsations have been detected from \objectname{PSR B1951+32} 
\citep{murthy_1951}, 
and reported from \objectname{PSR B0656+14} \citep{murthy_0656}, they do
not appear as detectable point sources in the 3EG catalog, so they are not
considered here. 

As seen in Table~2, the pulsars, AGN, and unidentified sources clearly
differ in their variability.  The variability of the six pulsars is 
about 11\%,
not significantly higher than the predicted systematic variability for a
constant source of 10\%. 

When the firmly identified quasars and BL Lac objects are considered separately,
a difference is clear.  On the average, the BL Lacs are much less variable
than the quasars.  This confirms the observation made by \citet{muk_agn}.

The ten  sources in the 3EG SNR source class are somewhat less variable
than the low-latitude
unidentified sources as a whole, and show significantly 
more variation than the pulsar class.  
Of course, it is by no means certain that these sources are supernova
remnants.  They are found in crowded areas of the sky along with
other types of objects characteristic of star-forming regions
\citep{ion_unid}. \citet{romero} identify 17 possible SNR
associations, of which 7 are also in the 3EG SNR list.
These two SNR collections have very similar variability properties.
If the gamma rays are produced in a large region of the SNR, then the flux
should be steady.  Observable variation would be an indication of gamma
production in a pulsar wind or perhaps of the presence of a background
blazar.

The change in unidentified source variability with latitude can be seen
in Table~2.  
It is clear from the spatial distribution of the unidentified
sources that there are multiple types of astrophysical objects which have not
been identified~\citep{grenier, gehrels, kanbach_characteristics}.  
It is clear that the
different types of unidentified sources have different variabilities as well.
The high latitude sources have a variability index consistent with that of the
identified AGN, and many of them have plausible AGN counterparts \citep{Mattox97,
mattox_id01,dse}.  
The low latitude unidentified sources exhibit some variability,
at a level a little more than the sources coincident with SNR. 

In the literature a number of other possible associations with source
classes have been suggested, not based on the classifications in the
3EG catalog. Table~2 also summarizes the properties of these classes.

\citet{grenier_texas02} listed 13 3EG sources with positions
consistent with radio pulsars and reasonable $L_\gamma/\dot{E}$ ratios.
Six of these (3EG~J1013$-$5915, 3EG~J1420$-$6038, 3EG~J1837$-$0423,
3EG~J1856+0114, 3EG~J2021+3716, 3EG~J2227+6122) have pulsar wind 
nebulae (PWN) which might be significant gamma sources.  (3EG~J1410$-$6147 is
not counted as a PWN source because it can be associated with
two pulsars, but the one with a PWN can be discounted on energetic
grounds.)  The seven non-PWN sources (3EG~J1014$-$5705,
3EG~J1102$-$6103, 3EG~J1410$-$6147, 3EG~J1639$-$4702,
3EG~J1714$-$3857, 3EG~J1824$-$1514, 3EG~J1837$-$0606) have an 
average $\delta$ of
$0.21 \pm 0.15$, while the six PWN sources have an average $\delta$
of $0.66 \pm 0.29$.  These numbers suggest that most of the
associations are correct and that PWNe can be strong, variable
gamma emitters.  Three of the PWN sources are among the
variable Galactic plane sources in \sect{plane}.

Four of the identified gamma-ray pulsars are known to have wind nebulae:
Crab, Vela, PSR B1046$-$58, and PSR B1706$-$44.  Their gamma
emission is clearly dominated by the pulsars.  Any variation in the
nebular output in the $> 100$ MeV band is hidden from this analysis.
The Crab Nebula, for instance, is variable on a scale of years
\citep{deJager1}, but the changes can be seen only in a certain energy 
band, and only in the off-pulse emission.
Note:  We do not consider the complex source
3EG~J0222+4253 here because of the confusion of emission from its
pulsar and blazar counterparts \citep{cat3}.

Three other 3EG sources have plausible associations with isolated
neutron stars.  They have hard spectra, large gamma to X-ray
flux ratios, and no known pulsar counterparts, so they might
be Geminga-class objects: 3EG~J0010+7309 \citep{brazier0010},
3EG~J1835+5918 \citep{reimer1835,chandler01,mirabal_1835},
and 3EG~J2020+4017 \citep{brazier2020}.  Individually they show
little sign of variability.  As a class, their average $\delta$
is $0.13 \pm 0.11$, quite consistent with the identified pulsars.

\citet{grenier2001} suggested an association of \egret\ sources with
Gould's Belt.  
\citet{perrot_gould} produced a model of the Belt based on a
3D simulation of Galactic dynamics.
There are 47 unidentified 3EG sources whose positions are consistent
with their model.  Of these, four are omitted from this analysis
as described in \sect{omissions}.  Ten more are at $|b|<5\arcdeg$,
so they are likely to be contaminated by the Galactic plane population(s).
The remaining 33 have an average $\delta$ of 0.49.  This 
is a moderately low value, but these sources as a class are definitely
more variable than the pulsars. Perhaps a fraction of them are
Myr-old pulsars \citep{hardingzhang}.

\citet{romero} made lists of possible associations with very young
objects: WR stars, Of stars,
and OB associations.  These all have an average $\delta$ of 0.26 to 0.45,
fairly steady but more variable than pulsars.

It has been proposed \citep{ClusterCoinc,ColaCluster} that some 
high-latitude \egret\ sources my be galaxy clusters which produce
gamma rays in their intracluster gas.  Of course such a source
would be non-variable on a scale of months to years. \citet{ClusterCoinc}
picked a sample of seven \egret\ sources, chosen on the basis of
steady emission and other criteria.  
3EG~J1310$-$0517 is omitted from our analysis, and the other six
show no significant evidence of variation.
The upper limits on $\delta$ are not tight (0.86 to 1.30),
so the non-variability of these sources is an open question.
\citet{ColaCluster} has a list of nine probable associations, 
three of which are also in the \citet{ClusterCoinc} list.    
However, he includes \objectname{3EG J0253$-$0345} and 
\objectname{3EG J0215+1123}.  Both of those are variable.

As described in \sect{PREV_METH}, \citet{gehrels} produced a list of
120 ``steady'' sources and \citet{grenier2001} a list of 88
``persistent'' sources.  Most of the ``persistent'' sources are
also in the the ``steady'' list.  These lists
were attempts to eliminate strongly flaring sources and ones which
are only marginally detectable.  Apparently they succeeded, since
the mean $\delta$ values (0.32 and 0.35) are fairly low.  
However, there are
several examples of clearly variable sources in these lists.

\subsection{The Lowest Variability Sources\label{lovsources}}

With the variability of the source classes in hand, it is possible to
see which sources can be excluded from a source class based on
the variability data.  We begin by examining those sources which
are inconsistent with being strongly variable: that is, the upper
bound ($\delta_\mathrm{max}$) of the 68\% region for $\delta$ is small.  
There are 8 sources for which
this upper bound is less than 0.4, significantly lower than
the typical value for AGN or unidentified sources.


Five of the six pulsars are in the list, with PSR B1055$-$52 barely 
excluded.  
The possibility that others on the list might also be pulsars is intriguing.  
The sources are discussed individually below.

The position of the \objectname{Crab} pulsar \objectname{3EG J0534+2200}, 
at the top of the list, is a bit 
surprising given its listing in \citet{mclaughlin} as moderately variable.
There are two reasons for this.  The first is the difference in the methods:
\citeauthor{mclaughlin} report the confidence level with which the hypothesis
of steady emission can be rejected.
In their analysis, the bright Crab, Geminga,
and Vela pulsars are all inconsistent with being constant at the 95\% level
due to systematic errors and the small statistical uncertainties.
The second reason is the difference in data used:  McLaughlin \etal\ used
all measurements up to to 30\deg\ from the instrument axis.  In this analysis,
we have used only data out to 25\deg, because of the larger systematic errors
at higher inclinations.

\objectname{3EG J1048$-$5840} is coincident with \objectname{PSR B1046$-$58}, 
a high $\dot{E}/d^2$ pulsar.
Although pulsed emission was not seen in the first three years of 
\egret\ data~\citep{joethesis},
there has been a detection in more recent data \citep{kaspi2000}.

\objectname{3EG J0222+4253} is coincident with the BL Lac \objectname{3C 66A}, 
and is 1\arcdeg\ from
the position of \objectname{PSR J0218+4232}, from which pulsed emission above 100 MeV 
has been reported \citep{verbunt, kuiper0218,kuiper2002}.  
In the 3EG catalog, the pulsar
was not detected as a separate source, and it is suggested that the
flux between 100 MeV and 1 GeV is primarily from the pulsar,
while the flux above 1 GeV is mostly
from 3C 66A \citep{cat3}.  The flux measurements of this ``source'' will be
dominated by the lower energy emission, hence the low variability.
\citet{wallace} found a significant variation of this ``source'' on
a 2-day timescale, likely from the AGN. Since our analysis uses only
full observations, we can't confirm or reject this variation.

The $\delta$ value of the Vela pulsar \objectname{3EG J0834$-$4511}
is slightly higher than the 10\% systematic variation.  
This is due to the presence
of several artifact sources nearby \citep{artifacts}.  
In a few viewing periods, the measured
flux of Vela is low, and the flux from an artifact is quite high.  
Because nearby source fluxes were not re-fit for each point on the 
likelihood-vs-intensity curve, such ``flux leaking'' can lead to an 
apparently high variability.  By fixing the fluxes of the spurious 
sources one could lower Vela's variability, but only if
those nearby sources were not themselves variable.  This was
not done because of the unknown biases that could result.

\objectname{3EG J2020+4017} is unidentified, but its position is coincident 
with the \objectname{$\gamma$-Cyg} supernova remnant, so it is thus
an intriguing pulsar candidate.  No pulsed signal has been detected from it,
however, and the possibility exists that it is a radio-quiet pulsar
\citep{brazier2020,chandler01}.

\objectname{3EG J1734$-$3232}, another unidentified source, is the last 
member of this group.

It is noteworthy that \objectname{3EG J0721+7120} almost qualifies for 
this list.
It is identified in the 3EG Catalog as the BL Lac object
\objectname{0716+714} \citep{cat3, cvm_agn, muk_agn}.
The upper end of the 68\% confidence region for
$\delta$ is 0.38, well below the typical AGN variability of $\sim 0.7$.
In large part, this is because this source had no dramatic flares during any
observations.
In addition, the viewing periods with the highest and lowest fluxes were 
each combined with less extreme observations by the one-month averaging 
system (\sect{method}, \sect{timescales}).  
A more sophisticated model might pick up longer or shorter term trends.

\subsection{The Highest Variability Sources}

From Table~1, 18 sources have a 68\% confidence region with a
lower bound greater than 0.7.  These include 11 sources identified as AGN.
Four of the unidentifed sources (\objectname{3EG J1227+4302}, 
\objectname{3EG J1607$-$1101}, \objectname{3EG J2006$-$2321}, and
\objectname{3EG J2251$-$1341}) are at high Galactic
latitudes, and are generally believed to be AGN~\citep{cat3, mattox_id01,
Wallace02}.  
The three remaining low latitude, unidentified sources are 
\objectname{3EG J1704$-$4732}, 
\objectname{3EG J1828+0142}, and
\objectname{3EG J1837$-$0423}.
\citet{halpern_blazar03} 
and \citet{dse} propose a blazar identification for
\objectname{3EG J1828+0142}.
\citet{Tavani97} discoverd the variability of 3EG J1837$-$0423,
but they reject a blazar identification.
None of the seven unidentified sources is classified as
steady or persistent.

Because the $V$ statistic used by \citet{mclaughlin} indicates how
improbable it is that the observations come from a constant source, the
use of the $\delta$ statistic in this way is nearly equivalent to the
use of $V$ to find variable sources.  Indeed, when there is a close
correspondence between 2EG and 3EG sources, sources with a high $V$
also have a high lower limit on $\delta$.

\subsection{Galactic Plane Sources\label{plane}}

\begin{figure}
\plotone{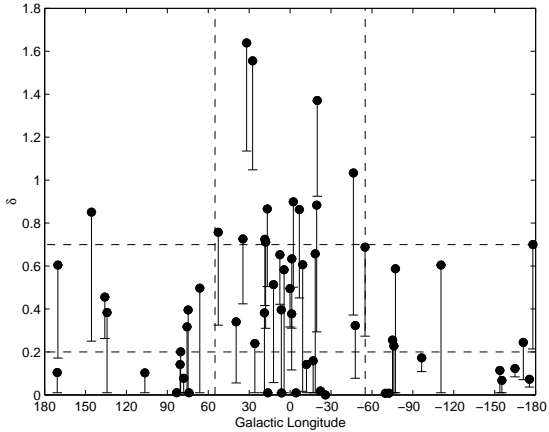}
\caption{\label{longitude}Values of $\delta$, a measure
of variability, for gamma sources within $\pm 6\arcdeg$ of the Galactic
Plane. For each source, the filled circle is the most likely value
of $\delta$ and the lower end of the bar is $\delta_\mathrm{min}$, 
the 68\% confidence lower limit.  The vertical 
dashed lines are drawn at longitudes of $\pm 55\arcdeg$
to guide the eye.  The horizontal lines are at the arbitrarily chosen
cutoff values for variable sources: 0.7 for the most likely $\delta$ and 0.2
for $\delta_\mathrm{min}$.  All of the highly variable sources are within
the central region of the Galaxy.}
\end{figure}

It is interesting to examine the behavior of the gamma sources near the
Galactic plane.  
There are many which show no sign of variability, but there
are some variable sources which are distributed non-uniformly in
Galactic longitude.  \fig{longitude} shows $\delta$ values
for the 66 sources within 6\arcdeg\ of the Galactic plane.  
If we define a variable source as one with a most likely $\delta$
value $> 0.7$ or a minimum $\delta > 0.2$, then it can be seen that almost
all of the 17 variable, low-latitude sources are clumped within $55\arcdeg$ of
the Galactic center.  All 17 are unidentified.

\begin{figure}
\plotone{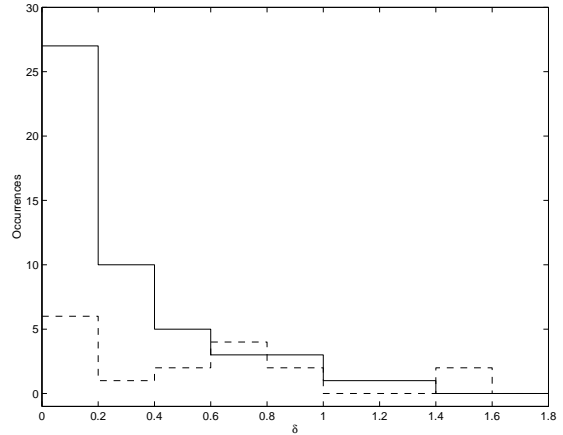}
\caption{\label{deltahist}Distributions of most likely $\delta$ values for
sources within 6\arcdeg\ of the Galactic plane.  Dashed line: Sources 
within 55\arcdeg\ longitude of the Galactic center. Solid line: Other sources.}
\end{figure}

\fig{deltahist} shows histograms of the most likely $\delta$ values for all
the sources within 6\arcdeg\ of the Galactic plane.
The dashed line shows the $\delta$ distribution for the sources with 
$|\ell| < 55\arcdeg$, while the solid line shows the others.
A Kolmogorov-Smirnov test \citep{eadie} rejects the hypothesis that these two 
sets of values are drawn from the same population with a p-value of 5.1\%.

\fig{longitude} shows values of $\delta$ for these sources as a function
of longitude.  The filled circles show the most likely $\delta$ values
and the ends of the bars show 68\%\ confidence lower limits
($\delta_\mathrm{min}$ in Table~1).  
The horizontal dashed lines are drawn at $\delta=0.7$ and 
$\delta_\mathrm{min}=0.2$ to guide the eye.

Some of the apparent variation might be due to flux leakage between
close companions, as with the Vela pulsar (\sect{lovsources}).
Among the 17 variable sources there are three pairs and a triple
within a few degrees.  No pair of these sources has correlated fluxes
with a correlation coefficient $> 0.18$, except \objectname{3EG J1823$-$1314}
and \objectname{3EG J1826$-$1302}.  Their correlation coefficient
is $-0.41$, with a p-value of 2.4\%.  The variation of these two
sources might be spurious, but the others seem to be on more solid ground.

The distribution of variable sources is consistent with a young Galactic
population, as traced by molecular clouds \citep{dameCO2001}, OB associations
\citep{kaaret}, SNRs and HII regions \citep{ion_unid,sturner_snr,sturner1996},
WR stars and Of stars \citep{romero}.
The concentration in the inner radian and their latitude
spread indicate typical distances of 5--8 kpc and luminosities
$L(E>$~100~MeV, $4\pi$~sr) of (1--25)~$\times\, 10^{35}$
(D/5~kpc)$^2$ erg~s$^{-1}$.
This is slightly higher than the range found by 
\citet{kanbach_characteristics} for all the low-latitude sources in 
the 2EG catalog, both variable and non-variable.

This luminosity range is typical of X-ray binaries
and 10 kyr-old pulsars.  The latter are expected to be steady
emitters, but variable emission from pulsar wind nebulae
is possible.  In particular,
plerionic activity is more likely to explain the
hard but variable 3EG~J1856+0114 source than pulsed emission
from the 20 kyr-old pulsar or cosmic-ray emission in the
surrounding W44 supernova shell.  The same may be true for
3EG~J1420$-$6038 in the Kookaburra nebula and 3EG~J1837$-$0423
near the G27.8+0.6 plerion.  3EG~J1809$-$2328 also coincides with
a pulsar wind nebula, but inverse-Compton emission from electrons
scattering the nebular synchrotron emission or the ambient radiation
field from the nearby OB association fail to explain the
source brightness \citep{braje}.  A 21 kyr-old pulsar,
PSR~J1826$-$1334, found near the edge of the 3EG~J1826$-$1302
error box, could power the gamma-ray source.  Pulsars are
also found near 3EG~J1308$-$6112 and 3EG~J1928+1733, but they
are much too weak to explain the sources.  

Variability over
months is common in accreting systems.  One high-mass X-ray
binary, GX~304$-$1, hosting a Be star and a neutron star on an eccentric
132.5-day orbit, is found toward 3EG~J1308$-$6112.  It is known
to exhibit strong long-term variability in X rays.
Electrons accelerated at the shock between the pulsar and stellar
winds have been proposed to shine up to TeV energies by upscattering
the stellar radiation \citep{tavar,kirk},
but the 272~s pulsar in GX~304$-$1 is not likely to power a
strong, energetic wind.  3EG~J1736$-$2908 coincides with the X-ray
source GRS~1734$-$292, which has been recently identified as a nearby
Seyfert 1 galaxy (Z = 0.0214) with radio jet emission and hard X-ray
emission up to 400 keV \citep{marti}.  Confirmation of variable
emission above 100 MeV from this galaxy would yield a first example
of the activity of a Seyfert jet at high energy and suggest a valuable
target for TeV telescopes. 

Curiously, 11 of the 17 variable sources are in the ``steady'' category,
and 10 are ``persistent.''  This happens because they are fairly bright.
As long as they don't vanish entirely for many observations, they
still meet the selection criteria.

\subsection{The Galactic Center}

The source nearest to the Galactic center, \objectname{3EG J1746$-$2851}, 
seems to be one of these variable sources.  
\citet{hanscenter} found only weak evidence that this source
might be variable.  With more data and a more sensitive method, 
the statistical evidence of its variability is fairly strong ($V_{12}=2.35$).
This would rule out any model in which this
object is a cluster of neutron stars, a condensation of dark matter particles,
or other non-compact object \citep{pohl97,markoff99,hanscenter,
roldan00,bertone02, CesariniGC}.

This conclusion is reinforced by \citet{hooper}, who show that the
position of the \egret\ source is not consistent with the
Galactic Center.  3EG J1746$-$2851 may be a field source like the
other 16 low-latitude variables.  The probablility of such a chance
positional coincidence near the Galactic Center is about 2.7\%.

Since the diffuse emission is most intense in the neighborhood 
of the Galactic Center, it is conceivable that errors in the background 
model or instrument PSF will have their greatest effect there.
To check this we correlated the flux of \objectname{3EG J1746$-$2851}
 with the angle 
to the instrument pointing direction.  The correlation coefficient
is 0.46, which has a p-value of 1.5\%.  If the maximum viewing
angle is reduced from $25^\circ$ to $17^\circ$, seven observations
are eliminated. The flux-angle correlation is reduced to an insignificant
level, but the estimate of $\delta$ becomes consistent with zero.

If there is a systematic problem at the Galactic Center, it should
show up also in measurements of nearby sources.  There are four
other sources within $6^\circ$ of the Center: \objectname{3EG~J1734$-$3232},
\objectname{3EG~J1736$-$2908}, \objectname{3EG J1741$-$2312}, and 
\objectname{3EG~J1744$-$3011}.
None of them shows such a large correlation of viewing angle
and flux, and in one case the correlation is negative.

However, 3EG~J1741$-$2321 and 3EG~J1744$-$3011 have a substantial 
negative flux correlation ($-0.33$ and $-0.42$ respectively)
with 3EG~J1746$-$2851.  This might be a sign of flux leakage,
although 3EG~J1741$-$2312 is rather far away. 

In view of these somewhat ambiguous signs of extra systematic
errors, we must declare that the variability of \objectname{3EG J1746$-$2851}
is only a tentative conclusion.

\section{Conclusions\label{CONCLUSIONS}}

It is possible to characterize the flux variation of \egret\ gamma-ray sources
between viewing periods in a consistent way by using a simple model for the
flux distribution and combining likelihood functions.  This method
produces useful results for almost all of the sources in the Third \egret\
catalog, using both positive detections and upper limits in a uniform manner.
The resulting table displays the average fractional variation $\delta$ and its
confidence interval.  For faint sources the confidence interval is necessarily
large, allowing a wide range of possible behaviors.

The identified rotation-powered pulsars all show a small variability, consistent
with the estimated 10\% systematic errors in the flux measurements.
In general, identified blazars are highly variable, although there are a
few exceptions.  The unidentified sources show a range of behaviors.
Their average variability increases with Galactic latitude, indicating that
there are multiple populations.  Unidentified sources associated with
pulsar with wind nebulae are, as a class, more variable than the ones 
associated with pulsars without nebulae.
The highest-latitude unidentified sources
have an average behavior similar to blazars.

There is a population of 17 highly-variable sources along the
Galactic plane.  These are concentrated in longitude within 1 radian of
the Galactic center.  This is probably a high-luminosity population of
sources in the inner spiral arms and molecular ring, while the remaining 
less-variable sources are closer and less luminous.

\acknowledgements
This work was supported by NASA grant NAG5-1605 and the ARCS Foundation.
The authors aknowledge helpful discussions with David Thompson,
Mallory Roberts, Olaf Reimer, Robert Hartman, and Brian Jones.
This research has made use of NASA's Astrophysics Data System and the
SIMBAD database, operated at CDS, Strasbourg, France.

\raggedright

\clearpage

\clearpage

\LongTables
\setcounter{table}{0}
\begin{deluxetable}{lrrrccrcll}
\tablecaption{Variability $\delta$ of sources in the Third \egret\ catalog}
\tablewidth{0pt}
\tablehead{
             \colhead{3EG Name} & \colhead{$\ell$} & \colhead{b} &
             \colhead{$\delta$} & \colhead{$\delta_\mathrm{min}$} & 
             \colhead{$\delta_\mathrm{max}$} &
             \colhead{$V_{12}$} & \colhead{Num} & \colhead{Classes} &
             \colhead{ID} \\
 & \colhead{(deg)} & \colhead{(deg)} & & & & & 
   \colhead{Obs} & & }
\startdata
\objectname{J0010+7309} &  119.92  & +10.54  & 0.26 & \phn\phd\phn0 & 0.73 & 0.40 & \phn3 & U5,SNR,Std,Pst,Gem & CTA 1 \\
\objectname{J0038$-$0949} &  112.69  & $-$72.44  & 0 & \phn\phd\phn0 & 1.34 & \nodata & \phn3 & U30,Std,Pst &   \\
\objectname{J0118+0248} &  136.23  & $-$59.36  & 1.18 & 0.48 & 2.18 & 1.64 & \phn4 & a & 0119+041 \\
\objectname{J0130$-$1758} &  169.71  & $-$77.11  & 0 & \phn\phd\phn0 & 0.63 & \nodata & \phn4 & a & 0130$-$171 \\
\objectname{J0159$-$3603} &  248.89  & $-$73.04  & 0 & \phn\phd\phn0 & 0.85 & \nodata & \phn3 & U30,Std,Pst &   \\
\objectname{J0204+1458} &  147.95  & $-$44.32  & 1.29 & 0.60 & 2.54 & 2.62 & \phn3 & QSO & 0202+149 \\
\objectname{J0210$-$5055} &  276.10  & $-$61.78  & 0.31 & 0.15 & 0.57 & 1.26 & \phn6 & QSO & 0208$-$512 \\
\objectname{J0215+1123} &  153.75  & $-$46.37  & 1.27 & 0.59 & 2.48 & 2.10 & \phn3 & U30 &   \\
\objectname{J0222+4253} &  140.14  & $-$16.77  & 0 & \phn\phd\phn0 & 0.37 & \nodata & \phn4 & BLL & 3C 66A, PSR J0218+4232 \\
\objectname{J0229+6151} &  134.20  & +1.15  & 0 & \phn\phd\phn0 & 0.51 & \nodata & \phn5 & U0,Of,OB,G,Std,Pst &   \\
\objectname{J0237+1635} &  156.77  & $-$39.11  & 0.89 & 0.46 & 1.89 & 5.07 & \phn3 & BLL & 0235+164 \\
\objectname{J0239+2815} &  150.21  & $-$28.80  & 0 & \phn\phd\phn0 & 0.41 & \nodata & \phn5 & a & 0234+285 \\
\objectname{J0241+6103} &  135.87  & +0.99  & 0.38 & 0.17 & 0.79 & 1.41 & \phn5 & U0,G,Std,Pst &   \\
\objectname{J0245+1758} &  157.62  & $-$37.11  & 1.14 & 0.44 & 2.16 & 1.38 & \phn4 & U30,Std &   \\
\objectname{J0253$-$0345} &  179.70  & $-$52.56  & 1.34 & 0.63 & 2.98 & 2.03 & \phn2 & U30 &   \\
\objectname{J0323+5122} &  145.64  & $-$4.67  & 0.92 & 0.31 & 1.68 & 1.26 & \phn7 & U0,G,Std &   \\
\objectname{J0329+2149} &  164.90  & $-$27.88  & 0 & \phn\phd\phn0 & 1.55 & \nodata & \phn5 & U15,Std &   \\
\objectname{J0340$-$0201} &  188.00  & $-$42.45  & 0 & \phn\phd\phn0 & 1.24 & \nodata & \phn2 & QSO & 0336$-$019 \\
\objectname{J0348+3510} &  159.06  & $-$15.01  & 0.45 & \phn\phd\phn0 & 1.29 & 0.24 & \phn6 & U15,G,Std,Pst &   \\
\objectname{J0348$-$5708} &  269.35  & $-$46.79  & 1.19 & \phn\phd\phn0 & 1.86 & 0.79 & \phn8 & U30 &   \\
\objectname{J0404+0700} &  184.00  & $-$32.15  & 0.39 & \phn\phd\phn0 & 1.53 & 0.11 & \phn6 & U30,Std,Pst &   \\
\objectname{J0407+1710} &  175.63  & $-$25.06  & 0.90 & \phn\phd\phn0 & 1.59 & 0.55 & \phn9 & U15,G,Std &   \\
\objectname{J0412$-$1853} &  213.90  & $-$43.29  & 0 & \phn\phd\phn0 & 2.10 & \nodata & \phn2 & QSO & 0414$-$189 \\
\objectname{J0416+3650} &  162.22  & $-$9.97  & 0.59 & 0.10 & 1.15 & 0.89 & \phn9 & a & 0415+379 \\
\objectname{J0422$-$0102} &  194.88  & $-$33.12  & 0 & \phn\phd\phn0 & 1.08 & \nodata & \phn4 & QSO & 0420$-$014 \\
\objectname{J0423+1707} &  178.48  & $-$22.14  & 0.42 & \phn\phd\phn0 & 1.02 & 0.42 & 10 & U15,G,Std,Pst &   \\
\objectname{J0426+1333} &  181.98  & $-$23.82  & 0 & \phn\phd\phn0 & 0.73 & \nodata & 12 & U15,G,Std,Pst &   \\
\objectname{J0429+0337} &  191.44  & $-$29.08  & 0 & \phn\phd\phn0 & 0.73 & \nodata & \phn8 & U15,Std,Pst &   \\
\objectname{J0433+2908} &  170.48  & $-$12.58  & 0.40 & 0.10 & 0.75 & 0.82 & 11 & QSO & 0430+2859 \\
\objectname{J0435+6137} &  146.50  & +9.50  & 0 & \phn\phd\phn0 & 0.54 & \nodata & \phn6 & U5,Std,Pst &   \\
\objectname{J0439+1105} &  186.14  & $-$22.87  & 0 & \phn\phd\phn0 & 0.74 & \nodata & 12 & U15,G,Std,Pst &   \\
\objectname{J0439+1555} &  181.98  & $-$19.98  & 1.08 & 0.48 & 1.63 & 1.20 & 12 & U15,G &   \\
\objectname{J0442$-$0033} &  197.20  & $-$28.46  & 1.59 & 1.12 & 2.31 & 14.59 & \phn8 & QSO & 0440$-$003 \\
\objectname{J0450+1105} &  187.86  & $-$20.62  & 1.13 & 0.78 & 1.64 & 8.42 & 12 & QSO & 0446+112 \\
\objectname{J0456$-$2338} &  223.96  & $-$34.98  & 0.21 & \phn\phd\phn0 & 1.42 & 0.05 & \phn5 & QSO & 0454$-$234 \\
\objectname{J0458$-$4635} &  252.40  & $-$38.40  & 0.41 & \phn\phd\phn0 & 1.19 & 0.23 & \phn6 & QSO & 0454$-$463 \\
\objectname{J0459+0544} &  193.99  & $-$21.66  & 0.74 & \phn\phd\phn0 & 1.43 & 0.54 & 10 & QSO & 0459+060 \\
\objectname{J0459+3352} &  170.25  & $-$5.77  & 0.59 & 0.14 & 1.18 & 0.95 & 12 & U5,G,Std,Pst &   \\
\objectname{J0500+2529} &  177.18  & $-$10.28  & 0 & \phn\phd\phn0 & 1.22 & \nodata & 12 & U5,G &   \\
\objectname{J0500$-$0159} &  201.35  & $-$25.47  & 1.09 & 0.58 & 1.71 & 2.29 & \phn9 & QSO & 0458$-$020 \\
\objectname{J0510+5545} &  153.99  & +9.42  & 0 & \phn\phd\phn0 & 0.41 & \nodata & \phn7 & U5,Std,Pst &   \\
\objectname{J0512$-$6150} &  271.25  & $-$35.28  & 0 & \phn\phd\phn0 & 0.71 & \nodata & \phn8 & a & 0506$-$612 \\
\objectname{J0520+2556} &  179.65  & $-$6.40  & 0 & \phn\phd\phn0 & 0.53 & \nodata & 12 & U5,G,Pst &   \\
\objectname{J0530+1323} &  191.37  & $-$11.01  & 0.74 & 0.55 & 1.05 & $\infty$ & 13 & QSO & 0528+134 \\
\objectname{J0530$-$3626} &  240.94  & $-$31.29  & 0.58 & \phn\phd\phn0 & 1.75 & 0.56 & \phn3 & a & 0521$-$365 \\
\objectname{J0531$-$2940} &  233.44  & $-$29.31  & 0.85 & \phn\phd\phn0 & 1.91 & 0.60 & \phn4 & a & 0537$-$286 \\
\objectname{J0533+4751} &  162.61  & +7.95  & 0 & \phn\phd\phn0 & 0.51 & \nodata & \phn8 & U5,Std,Pst &   \\
\objectname{J0533$-$6916} &  279.73  & $-$32.09  & 0.29 & \phn\phd\phn0 & 0.79 & 0.13 & \phn8 &  & LMC \\
\objectname{J0534+2200} &  184.56  & $-$5.78  & 0.08 & 0.03 & 0.15 & \nodata & 12 & PSR & Crab \\
\objectname{J0540$-$4402} &  250.08  & $-$31.09  & 0.75 & 0.46 & 1.27 & 6.18 & \phn6 & BLL & 0537$-$441 \\
\objectname{J0542+2610} &  182.02  & $-$1.99  & 0.57 & \phn\phd\phn0 & 1.17 & 0.76 & 12 & U0,SNR,RSN & S147 \\
\objectname{J0542$-$0655} &  211.28  & $-$18.52  & 1.38 & 0.84 & 2.14 & 1.93 & \phn7 & a & 0539$-$057 \\
\objectname{J0546+3948} &  170.75  & +5.74  & 0.10 & \phn\phd\phn0 & 0.68 & \nodata & 11 & U5,Std,Pst &   \\
\objectname{J0613+4201} &  171.32  & +11.40  & 0.72 & 0.09 & 1.38 & 0.87 & 10 & U5,Std,Pst &   \\
\objectname{J0616$-$0720} &  215.58  & $-$11.06  & 0 & \phn\phd\phn0 & 1.32 & \nodata & \phn8 & U5,G,Std &   \\
\objectname{J0616$-$3310} &  240.35  & $-$21.24  & 0.17 & \phn\phd\phn0 & 1.04 & 0.04 & \phn3 & U15,G,Std,Pst &   \\
\objectname{J0617+2238} &  189.00  & +3.05  & 0.25 & 0.10 & 0.46 & 0.82 & 12 & U0,SNR,RSN,Std,Pst & IC 443 \\
\objectname{J0622$-$1139} &  220.16  & $-$11.69  & 1.04 & 0.26 & 1.96 & 1.08 & \phn5 & QSO & 0616$-$116 \\
\objectname{J0631+0642} &  204.71  & $-$1.30  & 0.01 & \phn\phd\phn0 & 1.16 & \nodata & 13 & U0,RSN,Std &   \\
\objectname{J0633+1751} &  195.13  & +4.27  & 0.10 & 0.06 & 0.16 & \nodata & 10 & PSR & Geminga \\
\objectname{J0634+0521} &  206.18  & $-$1.41  & 0.01 & \phn\phd\phn0 & 1.12 & \nodata & 13 & U0,Of,OB,Std &   \\
\objectname{J0702$-$6212} &  272.65  & $-$22.56  & 1.01 & 0.34 & 1.65 & 1.06 & \phn9 & U15,Std &   \\
\objectname{J0706$-$3837} &  249.57  & $-$13.76  & 0 & \phn\phd\phn0 & 1.75 & \nodata & \phn4 & U5,G &   \\
\objectname{J0721+7120} &  143.98  & +28.02  & 0 & \phn\phd\phn0 & 0.41 & \nodata & \phn5 & BLL & 0716+714 \\
\objectname{J0724$-$4713} &  259.00  & $-$14.38  & 1.16 & 0.57 & 1.94 & 2.02 & \phn6 & U5,G,Std &   \\
\objectname{J0725$-$5140} &  263.29  & $-$16.02  & 0.89 & 0.29 & 1.64 & 1.18 & \phn7 & U15,G,Std,Pst &   \\
\objectname{J0737+1721} &  201.85  & +18.07  & 0 & \phn\phd\phn0 & 0.70 & \nodata & \phn3 & BLL & 0735+178 \\
\objectname{J0743+5447} &  162.99  & +29.19  & 1.21 & 0.67 & 2.06 & 4.78 & \phn5 & QSO & 0738+5451 \\
\objectname{J0747$-$3412} &  249.35  & $-$4.48  & 0.59 & \phn\phd\phn0 & 1.55 & 0.29 & \phn6 & U0,WR,G,Std &   \\
\objectname{J0808+4844} &  170.46  & +32.48  & 0 & \phn\phd\phn0 & 0.65 & \nodata & \phn5 & a & 0804+499?, 0809+483? \\
\objectname{J0808+5114} &  167.51  & +32.66  & 0 & \phn\phd\phn0 & 1.18 & \nodata & \phn4 & a & 0803+5126 \\
\objectname{J0808$-$5344} &  268.24  & $-$11.20  & 1.04 & 0.55 & 1.68 & 2.16 & \phn8 & U5,G,Std &   \\
\objectname{J0812$-$0646} &  228.64  & +14.62  & 0 & \phn\phd\phn0 & 0.90 & \nodata & \phn2 & a & 0805$-$077 \\
\objectname{J0821$-$5814} &  273.10  & $-$12.04  & 1.16 & 0.46 & 1.84 & 1.05 & \phn8 & U5,G &   \\
\objectname{J0828+0508} &  219.60  & +23.82  & 0 & \phn\phd\phn0 & 2.17 & \nodata & \phn2 & BLL & 0829+046 \\
\objectname{J0834$-$4511} &  263.55  & $-$2.79  & 0.16 & 0.10 & 0.28 & 0.61 & \phn7 & PSR & Vela  \\
\objectname{J0845+7049} &  143.54  & +34.43  & 0.62 & 0.23 & 1.43 & 1.38 & \phn4 & QSO & 0836+710 \\
\objectname{J0852$-$1216} &  239.06  & +19.99  & 1.21 & 0.52 & 2.26 & 1.94 & \phn4 & QSO & PMN J0850$-$1213 \\
\objectname{J0853+1941} &  206.81  & +35.82  & 0 & \phn\phd\phn0 & 0.81 & \nodata & \phn3 & BLL & 0851+202 \\
\objectname{J0903$-$3531} &  259.40  & +7.40  & 0.52 & \phn\phd\phn0 & 1.28 & 0.70 & \phn7 & U5,Std &   \\
\objectname{J0910+6556} &  148.30  & +38.56  & 0.64 & \phn\phd\phn0 & 1.61 & 0.60 & \phn4 & U30,Std &   \\
\objectname{J0917+4427} &  176.11  & +44.19  & 0 & \phn\phd\phn0 & 0.53 & \nodata & \phn5 & a & 0917+449 \\
\objectname{J0952+5501} &  159.55  & +47.33  & 0.39 & \phn\phd\phn0 & 0.88 & 0.31 & \phn9 & QSO & 0954+556 \\
\objectname{J0958+6533} &  145.75  & +43.13  & 0.96 & 0.42 & 1.72 & 2.29 & \phn6 & BLL & 0954+658 \\
\objectname{J1009+4855} &  166.87  & +51.99  & 0 & \phn\phd\phn0 & 0.94 & \nodata & \phn9 & a & 1011+496 \\
\objectname{J1013$-$5915} &  283.93  & $-$2.34  & 0.14 & \phn\phd\phn0 & 0.52 & 0.18 & \phn9 & U0,G,RSN,Std,Pst,PWN &   \\
\objectname{J1014$-$5705} &  282.80  & $-$0.51  & 0 & \phn\phd\phn0 & 0.55 & \nodata & 10 & U0,G,Std,Pst,noPWN &   \\
\objectname{J1027$-$5817} &  284.94  & $-$0.52  & 0.21 & \phn\phd\phn0 & 0.46 & 0.39 & \phn9 & U0,OB,G,Std,Pst &   \\
\objectname{J1045$-$7630} &  295.66  & $-$15.45  & 0 & \phn\phd\phn0 & 0.44 & \nodata & \phn7 & U15,Std,Pst &   \\
\objectname{J1048$-$5840} &  287.53  & +0.47  & 0 & \phn\phd\phn0 & 0.31 & \nodata & \phn8 & PSR,Std & PSR B1046$-$58 \\
\objectname{J1052+5718} &  149.47  & +53.27  & 0.24 & \phn\phd\phn0 & 1.02 & 0.10 & \phn6 & a & 1055+567 \\
\objectname{J1058$-$5234} &  285.98  & +6.65  & 0 & \phn\phd\phn0 & 0.47 & \nodata & \phn8 & PSR & PSR B1055$-$52 \\
\objectname{J1102$-$6103} &  290.12  & $-$0.92  & 0 & \phn\phd\phn0 & 1.08 & \nodata & \phn7 & U0,WR,OB,G,SNR,Std,Pst,noPWN & MSH 11$-$62 \\
\objectname{J1104+3809} &  179.83  & +65.03  & 0.35 & \phn\phd\phn0 & 0.80 & 0.46 & \phn7 & BLL & Mrk 421 \\
\objectname{J1133+0033} &  264.52  & +57.48  & 0.82 & \phn\phd\phn0 & 1.51 & 0.65 & \phn8 & U30 &   \\
\objectname{J1134$-$1530} &  277.04  & +43.48  & 1.12 & 0.61 & 1.84 & 3.79 & \phn7 & a & 1127$-$145 \\
\objectname{J1200+2847} &  199.42  & +78.38  & 1.17 & 0.75 & 1.77 & 5.49 & \phn9 & QSO & 1156+295 \\
\objectname{J1212+2304} &  235.57  & +80.32  & 0 & \phn\phd\phn0 & 1.13 & \nodata & 11 & U30 &   \\
\objectname{J1219$-$1520} &  291.56  & +46.82  & 1.10 & 0.56 & 1.72 & 1.77 & \phn9 & U30 &   \\
\objectname{J1222+2315} &  241.87  & +82.39  & 1.09 & 0.60 & 1.65 & 1.61 & 10 & U30 &   \\
\objectname{J1222+2841} &  197.27  & +83.52  & 0.62 & 0.22 & 1.20 & 1.17 & \phn8 & BLL & 1219+285 \\
\objectname{J1224+2118} &  255.07  & +81.66  & 0.45 & \phn\phd\phn0 & 0.98 & 0.62 & \phn9 & QSO & 1222+216 \\
\objectname{J1227+4302} &  138.63  & +73.33  & 1.36 & 0.76 & 2.37 & 2.34 & \phn4 & U30 &   \\
\objectname{J1229+0210} &  289.95  & +64.36  & 0.46 & 0.22 & 0.85 & 1.56 & \phn8 & QSO & 3C 273 \\
\objectname{J1230$-$0247} &  292.58  & +59.66  & 0.59 & \phn\phd\phn0 & 1.31 & 0.71 & \phn8 & QSO & 1229$-$021 \\
\objectname{J1234$-$1318} &  296.43  & +49.34  & 0.61 & \phn\phd\phn0 & 1.33 & 0.77 & \phn8 & U30,Std,Pst &   \\
\objectname{J1235+0233} &  293.28  & +65.13  & 0.21 & \phn\phd\phn0 & 0.86 & 0.11 & \phn8 & U30,Std,Pst &   \\
\objectname{J1236+0457} &  292.59  & +67.52  & 0.53 & \phn\phd\phn0 & 1.42 & 0.10 & \phn8 & a & 1237+0459 \\
\objectname{J1246$-$0651} &  300.96  & +55.99  & 0.60 & 0.11 & 1.23 & 0.89 & \phn8 & QSO & 1243$-$072 \\
\objectname{J1249$-$8330} &  302.86  & $-$20.63  & 0.71 & \phn\phd\phn0 & 1.64 & 0.62 & \phn6 & U15,Std &   \\
\objectname{J1255$-$0549} &  305.09  & +57.06  & 0.90 & 0.65 & 1.35 & $\infty$ & \phn8 & QSO & 3C 279 \\
\objectname{J1300$-$4406} &  304.60  & +18.74  & 0.54 & \phn\phd\phn0 & 1.40 & 0.30 & \phn8 & U15,Std &   \\
\objectname{J1308+8744} &  122.74  & +29.38  & 0 & \phn\phd\phn0 & 0.94 & \nodata & \phn6 & U15,Std &   \\
\objectname{J1308$-$6112} &  305.01  & +1.59  & 0.72 & 0.29 & 1.45 & 1.37 & \phn8 & U0,OB,G,Std &   \\
\objectname{J1314$-$3431} &  308.21  & +28.12  & 0 & \phn\phd\phn0 & 0.42 & \nodata & \phn6 & a & 1313$-$333 \\
\objectname{J1316$-$5244} &  306.85  & +9.93  & 0.42 & \phn\phd\phn0 & 0.98 & 0.49 & \phn8 & U5,G,Std,Pst &   \\
\objectname{J1323+2200} &  359.33  & +81.15  & 1.09 & 0.52 & 1.79 & 1.61 & \phn7 & a & 1324+224 \\
\objectname{J1324$-$4314} &  309.32  & +19.21  & 0 & \phn\phd\phn0 & 0.49 & \nodata & \phn8 &  & Cen A \\
\objectname{J1329+1708} &  346.29  & +76.68  & 0.62 & \phn\phd\phn0 & 1.36 & 0.14 & \phn9 & QSO & 1331+170 \\
\objectname{J1329$-$4602} &  309.83  & +16.32  & 0 & \phn\phd\phn0 & 0.76 & \nodata & \phn8 & U15,Std,Pst &   \\
\objectname{J1337+5029} &  105.40  & +65.04  & 0.53 & \phn\phd\phn0 & 1.30 & 0.41 & \phn5 & U30,Std,Pst &   \\
\objectname{J1339$-$1419} &  320.07  & +46.95  & 0.81 & \phn\phd\phn0 & 1.56 & 0.34 & \phn7 & QSO & 1334$-$127 \\
\objectname{J1347+2932} &   47.32  & +77.50  & 0.64 & \phn\phd\phn0 & 1.60 & 0.40 & \phn6 & U30,Std &   \\
\objectname{J1409$-$0745} &  333.88  & +50.28  & 1.42 & 1.00 & 2.07 & $\infty$ & \phn8 & QSO & 1406$-$076 \\
\objectname{J1410$-$6147} &  312.18  & $-$0.35  & 0.32 & 0.08 & 0.64 & 0.76 & \phn8 & U0,Of,OB,G,SNR,RSN,Std,Pst,noPWN & G312.4$-$0.4 \\
\objectname{J1420$-$6038} &  313.63  & +0.37  & 1.03 & 0.37 & 1.83 & 1.59 & \phn8 & U0,OB,G,Std,Pst,PWN &   \\
\objectname{J1429$-$4217} &  321.45  & +17.27  & 0.93 & 0.44 & 1.58 & 2.15 & \phn9 & QSO & 1424$-$418 \\
\objectname{J1447$-$3936} &  326.12  & +17.96  & 0.28 & \phn\phd\phn0 & 1.07 & 0.09 & \phn9 & U15,G,Std,Pst &   \\
\objectname{J1457$-$1903} &  339.88  & +34.60  & 0.49 & \phn\phd\phn0 & 1.65 & 0.07 & \phn6 & U30,Std &   \\
\objectname{J1500$-$3509} &  330.91  & +20.45  & 0 & \phn\phd\phn0 & 0.92 & \nodata & \phn9 & U15,G,Std,Pst &   \\
\objectname{J1504$-$1537} &  344.04  & +36.38  & 1.24 & 0.58 & 2.15 & 1.79 & \phn5 & a & 1504$-$166 \\
\objectname{J1512$-$0849} &  351.49  & +40.37  & 0 & \phn\phd\phn0 & 0.52 & \nodata & \phn5 & QSO & 1510$-$089 \\
\objectname{J1517$-$2538} &  339.76  & +26.60  & 0 & \phn\phd\phn0 & 1.59 & \nodata & \phn6 & a & 1514$-$241 \\
\objectname{J1527$-$2358} &  342.97  & +26.50  & 0.92 & \phn\phd\phn0 & 1.71 & 0.07 & \phn9 & U15 &   \\
\objectname{J1600$-$0351} &    6.30  & +34.81  & 0.01 & \phn\phd\phn0 & 1.62 & \nodata & \phn4 & U30 &   \\
\objectname{J1605+1553} &   29.18  & +43.84  & 0 & \phn\phd\phn0 & 0.70 & \nodata & \phn2 & BLL & 1604+159 \\
\objectname{J1607$-$1101} &    0.91  & +29.05  & 1.45 & 0.93 & 2.10 & 0.93 & 10 & U15 &   \\
\objectname{J1608+1055} &   23.03  & +40.79  & 1.09 & 0.25 & 2.72 & 1.15 & \phn2 & QSO & 1606+106 \\
\objectname{J1612$-$2618} &  349.76  & +17.64  & 1.04 & 0.48 & 1.62 & 1.47 & 12 & U15,G &   \\
\objectname{J1614+3424} &   55.15  & +46.38  & 0 & \phn\phd\phn0 & 0.41 & \nodata & \phn2 & QSO & 1611+343 \\
\objectname{J1616$-$2221} &  353.00  & +20.03  & 0 & \phn\phd\phn0 & 0.54 & \nodata & 12 & U15,G,Std,Pst &   \\
\objectname{J1621+8203} &  115.53  & +31.77  & 0 & \phn\phd\phn0 & 0.52 & \nodata & \phn4 & U30,Std &   \\
\objectname{J1625$-$2955} &  348.82  & +13.32  & 1.62 & 1.22 & 2.19 & $\infty$ & 12 & QSO & 1622$-$297 \\
\objectname{J1626$-$2519} &  352.14  & +16.32  & 1.05 & 0.54 & 1.60 & 2.03 & 13 & QSO & 1622$-$253 \\
\objectname{J1627$-$2419} &  353.36  & +16.71  & 0 & \phn\phd\phn0 & 0.57 & \nodata & 12 & U15,G,Std,Pst &   \\
\objectname{J1631$-$1018} &    5.55  & +24.94  & 0.35 & \phn\phd\phn0 & 1.01 & 0.21 & 12 & U15,G,Std,Pst &   \\
\objectname{J1631$-$4033} &  341.61  & +5.24  & 0.31 & \phn\phd\phn0 & 0.90 & 0.16 & 14 & U5,G,Std &   \\
\objectname{J1633$-$3216} &  348.10  & +10.48  & 0.59 & \phn\phd\phn0 & 1.25 & 0.50 & 13 & U5,G,Std &   \\
\objectname{J1634$-$1434} &    2.33  & +21.78  & 0 & \phn\phd\phn0 & 0.88 & \nodata & 12 & U15,G,Std,Pst &   \\
\objectname{J1635+3813} &   61.09  & +42.34  & 0 & \phn\phd\phn0 & 0.55 & \nodata & \phn2 & QSO & 1633+382 \\
\objectname{J1635$-$1751} &  359.72  & +19.56  & 0.83 & \phn\phd\phn0 & 1.48 & 0.16 & 13 & U15,G &   \\
\objectname{J1638$-$2749} &  352.25  & +12.59  & 0.47 & 0.17 & 0.88 & 1.06 & 14 & U5,G,Std,Pst &   \\
\objectname{J1638$-$5155} &  334.05  & $-$3.34  & 0.19 & \phn\phd\phn0 & 0.91 & 0.13 & 13 & U0,Std,Pst &   \\
\objectname{J1639$-$4702} &  337.75  & $-$0.15  & 0 & \phn\phd\phn0 & 0.57 & \nodata & 12 & U0,OB,RSN,Std,Pst,noPWN &   \\
\objectname{J1646$-$0704} &   10.85  & +23.69  & 0.75 & \phn\phd\phn0 & 1.41 & 0.76 & 12 & U15,G,Std,Pst &   \\
\objectname{J1649$-$1611} &    3.35  & +17.80  & 0 & \phn\phd\phn0 & 1.21 & \nodata & 14 & U15,G,Std,Pst &   \\
\objectname{J1652$-$0223} &   15.99  & +25.05  & 0 & \phn\phd\phn0 & 0.96 & \nodata & \phn6 & U15,G,Std,Pst &   \\
\objectname{J1653$-$2133} &  359.49  & +13.81  & 0.95 & \phn\phd\phn0 & 1.51 & 0.73 & 14 & U5,G &   \\
\objectname{J1655$-$4554} &  340.48  & $-$1.61  & 0.78 & 0.18 & 1.50 & 0.97 & 13 & U0,WR,OB,Std,Pst &   \\
\objectname{J1659$-$6251} &  327.32  & $-$12.47  & 0.73 & \phn\phd\phn0 & 1.66 & 0.47 & \phn6 & U5,Std &   \\
\objectname{J1704$-$4732} &  340.10  & $-$3.79  & 1.37 & 0.91 & 1.93 & 2.67 & 13 & U0 &   \\
\objectname{J1709$-$0828} &   12.86  & +18.25  & 0.67 & \phn\phd\phn0 & 1.40 & 0.37 & 11 & U15,G,Std,Pst &   \\
\objectname{J1710$-$4439} &  343.10  & $-$2.69  & 0.07 & \phn\phd\phn0 & 0.21 & \nodata & 14 & PSR & PSR B1706$-$44 \\
\objectname{J1714$-$3857} &  348.04  & $-$0.09  & 0.17 & \phn\phd\phn0 & 0.52 & 0.29 & 13 & U0,RSN,Std,Pst,noPWN &   \\
\objectname{J1717$-$2737} &  357.67  & +5.95  & 0.91 & 0.50 & 1.43 & 2.72 & 15 & U5,Std,Pst &   \\
\objectname{J1718$-$3313} &  353.20  & +2.56  & 0.82 & 0.43 & 1.36 & 2.50 & 15 & U0,OB &   \\
\objectname{J1719$-$0430} &   17.80  & +18.17  & 0 & \phn\phd\phn0 & 0.81 & \nodata & \phn8 & U15,G,Std,Pst &   \\
\objectname{J1720$-$7820} &  314.56  & $-$22.17  & 0.64 & \phn\phd\phn0 & 2.13 & 0.02 & \phn3 & a & 1716$-$771 \\
\objectname{J1726$-$0807} &   15.52  & +14.77  & 0.45 & \phn\phd\phn0 & 1.03 & 0.36 & 12 & U5,G,Std,Pst &   \\
\objectname{J1727+0429} &   27.27  & +20.62  & 0 & \phn\phd\phn0 & 0.74 & \nodata & \phn3 & QSO & 1725+044 \\
\objectname{J1733+6017} &   89.12  & +32.94  & 0.29 & \phn\phd\phn0 & 1.32 & 0.15 & \phn4 & U30,Std &   \\
\objectname{J1733$-$1313} &   12.03  & +10.81  & 0.37 & 0.14 & 0.68 & 1.01 & 12 & QSO & 1730$-$130 \\
\objectname{J1734$-$3232} &  355.64  & +0.15  & 0 & \phn\phd\phn0 & 0.37 & \nodata & 16 & U0,OB,RSN,Std,Pst &   \\
\objectname{J1735$-$1500} &   10.73  & +9.22  & 0.87 & \phn\phd\phn0 & 1.45 & 0.38 & 14 & U5,G &   \\
\objectname{J1736$-$2908} &  358.79  & +1.56  & 0.64 & 0.31 & 1.12 & 1.67 & 15 & U0,Std,Pst &   \\
\objectname{J1738+5203} &   79.37  & +32.05  & 0.51 & 0.10 & 1.37 & 0.85 & \phn3 & QSO & 1739+522 \\
\objectname{J1741$-$2050} &    6.44  & +5.00  & 0 & \phn\phd\phn0 & 0.57 & \nodata & 14 & U5,Std,Pst &   \\
\objectname{J1741$-$2312} &    4.42  & +3.76  & 0.26 & \phn\phd\phn0 & 0.78 & 0.15 & 14 & U0,Std,Pst &   \\
\objectname{J1744$-$0310} &   22.19  & +13.42  & 0.59 & \phn\phd\phn0 & 1.49 & 0.52 & \phn7 & QSO & 1741$-$038 \\
\objectname{J1744$-$3011} &  358.85  & $-$0.52  & 0.30 & \phn\phd\phn0 & 0.61 & 0.64 & 15 & U0,RSN,Std,Pst &   \\
\objectname{J1744$-$3934} &  350.81  & $-$5.38  & 0.60 & 0.04 & 1.14 & 0.82 & 12 & U5,Std,Pst &   \\
\objectname{J1746$-$1001} &   16.34  & +9.64  & 0.42 & \phn\phd\phn0 & 0.88 & 0.54 & 12 & U5,G,Std,Pst &   \\
\objectname{J1746$-$2851}\tablenotemark{a} &    0.11  & $-$0.04  & 0.48 & 0.29 & 0.75 & 2.35 & 15 & U0,RSN,Std,Pst &   \\
\objectname{J1757$-$0711} &   20.30  & +8.47  & 0.51 & \phn\phd\phn0 & 1.11 & 0.42 & 14 & U5,G,Std,Pst &   \\
\objectname{J1800$-$0146} &   25.49  & +10.39  & 0 & \phn\phd\phn0 & 0.58 & \nodata & \phn8 & U5,G,Std,Pst &   \\
\objectname{J1800$-$2338} &    6.25  & $-$0.18  & 0.16 & \phn\phd\phn0 & 0.58 & 0.26 & 16 & U0,SNR,RSN,Std,Pst & W28 \\
\objectname{J1800$-$3955} &  352.45  & $-$8.43  & 0 & \phn\phd\phn0 & 1.13 & \nodata & 13 & QSO & 1759$-$396 \\
\objectname{J1806$-$5005} &  343.29  & $-$13.76  & 0.88 & \phn\phd\phn0 & 1.56 & 0.34 & 10 & a & PMN J1808$-$5011 \\
\objectname{J1809$-$2328} &    7.47  & $-$1.99  & 0.71 & 0.46 & 1.13 & 3.93 & 15 & U0,OB,Std,Pst &   \\
\objectname{J1810$-$1032} &   18.81  & +4.23  & 0 & \phn\phd\phn0 & 0.58 & \nodata & 14 & U0,Std &   \\
\objectname{J1812$-$1316} &   16.70  & +2.39  & 0.82 & 0.47 & 1.35 & 2.73 & 15 & U0,Std,Pst &   \\
\objectname{J1813$-$6419} &  330.04  & $-$20.32  & 0 & \phn\phd\phn0 & 0.97 & \nodata & \phn4 & U15,Std,Pst &   \\
\objectname{J1822+1641} &   44.84  & +13.84  & 1.11 & 0.54 & 2.10 & 2.26 & \phn4 & U5,G &   \\
\objectname{J1823$-$1314} &   17.94  & +0.14  & 0.60 & 0.21 & 1.18 & 1.18 & 14 & U0,OB,SNR,Pst & Kes 67 \\
\objectname{J1824+3441} &   62.49  & +20.14  & 0.01 & \phn\phd\phn0 & 1.30 & \nodata & \phn8 & U15,G &   \\
\objectname{J1824$-$1514} &   16.37  & $-$1.16  & 0 & \phn\phd\phn0 & 0.84 & \nodata & 14 & U0,OB,RSN,Std,Pst,noPWN &   \\
\objectname{J1825+2854} &   56.79  & +18.03  & 0 & \phn\phd\phn0 & 1.33 & \nodata & \phn7 & U15,G &   \\
\objectname{J1825$-$7926} &  314.56  & $-$25.44  & 0.76 & \phn\phd\phn0 & 1.77 & 0.78 & \phn4 & U15,Std,Pst &   \\
\objectname{J1826$-$1302} &   18.47  & $-$0.44  & 0.88 & 0.50 & 1.45 & 3.22 & 14 & U0,OB,Std,Pst &   \\
\objectname{J1828+0142} &   31.90  & +5.78  & 1.60 & 1.02 & 2.48 & 6.89 & \phn6 & U5 &   \\
\objectname{J1832$-$2110} &   12.17  & $-$5.71  & 0.62 & 0.19 & 1.20 & 1.05 & 13 & QSO & 1830$-$210 \\
\objectname{J1834$-$2803} &    5.92  & $-$8.97  & 0 & \phn\phd\phn0 & 0.59 & \nodata & 14 & U5,Std,Pst &   \\
\objectname{J1835+5918} &   88.74  & +25.07  & 0.15 & \phn\phd\phn0 & 0.47 & 0.09 & \phn4 & U15,Std,Pst,Gem &   \\
\objectname{J1836$-$4933} &  345.93  & $-$18.26  & 0 & \phn\phd\phn0 & 0.97 & \nodata & \phn9 & U15,Std,Pst &   \\
\objectname{J1837$-$0423} &   27.44  & +1.06  & 1.50 & 0.97 & 2.26 & 2.95 & \phn8 & U0,RSN,PWN &   \\
\objectname{J1837$-$0606} &   25.86  & +0.40  & 0 & \phn\phd\phn0 & 0.67 & \nodata & \phn8 & U0,Std,Pst,noPWN &   \\
\objectname{J1847$-$3219} &    3.21  & $-$13.37  & 0.75 & \phn\phd\phn0 & 1.36 & 0.71 & 14 & U5,Std &   \\
\objectname{J1850+5903} &   88.92  & +23.18  & 0 & \phn\phd\phn0 & 1.58 & \nodata & \phn5 & U15 &   \\
\objectname{J1850$-$2652} &    8.73  & $-$11.76  & 0.81 & \phn\phd\phn0 & 1.42 & 0.59 & 13 & U5 &   \\
\objectname{J1856+0114} &   34.60  & $-$0.54  & 0.71 & 0.28 & 1.53 & 1.57 & \phn6 & U0,SNR,RSN,Std,Pst,PWN & W44 \\
\objectname{J1858$-$2137} &   14.21  & $-$11.15  & 0 & \phn\phd\phn0 & 0.80 & \nodata & 12 & U5,Std,Pst &   \\
\objectname{J1903+0550} &   39.52  & $-$0.05  & 0 & \phn\phd\phn0 & 0.42 & \nodata & \phn5 & U0,SNR,RSN,Std,Pst & G40.5$-$0.5 \\
\objectname{J1904$-$1124} &   24.22  & $-$8.12  & 0.08 & \phn\phd\phn0 & 0.76 & \nodata & 10 & U5,Std,Pst &   \\
\objectname{J1911$-$2000} &   16.87  & $-$13.22  & 0.30 & \phn\phd\phn0 & 0.82 & 0.25 & 11 & QSO & 1908$-$201 \\
\objectname{J1921$-$2015} &   17.81  & $-$15.60  & 1.24 & 0.80 & 1.76 & 2.19 & 12 & a & 1920$-$211 \\
\objectname{J1928+1733} &   52.71  & +0.07  & 0.76 & 0.33 & 1.57 & 1.83 & \phn6 & U0 &   \\
\objectname{J1935$-$4022} &  358.65  & $-$25.23  & 1.34 & 0.83 & 2.12 & 4.41 & \phn6 & QSO & 1933$-$400 \\
\objectname{J1937$-$1529} &   23.95  & $-$17.12  & 1.35 & 0.72 & 2.05 & 1.08 & \phn8 & QSO & 1936$-$155 \\
\objectname{J1940$-$0121} &   37.41  & $-$11.62  & 1.28 & 0.70 & 2.10 & 2.48 & \phn6 & U5 &   \\
\objectname{J1949$-$3456} &    5.25  & $-$26.29  & 1.17 & 0.64 & 1.84 & 2.59 & \phn8 & U15 &   \\
\objectname{J1955$-$1414} &   27.01  & $-$20.56  & 0.84 & 0.10 & 1.62 & 0.88 & \phn7 & U15,Std &   \\
\objectname{J1958+2909} &   66.23  & $-$0.16  & 0.32 & \phn\phd\phn0 & 0.86 & 0.31 & \phn9 & U0,Std,Pst &   \\
\objectname{J1958$-$4443} &  354.85  & $-$30.13  & 0 & \phn\phd\phn0 & 1.64 & \nodata & \phn4 & U30 &   \\
\objectname{J1959+6342} &   96.61  & +17.10  & 0 & \phn\phd\phn0 & 0.76 & \nodata & \phn6 & U15,Std,Pst &   \\
\objectname{J2006$-$2321} &   18.82  & $-$26.26  & 1.31 & 0.73 & 2.12 & 2.63 & \phn6 & U15 &   \\
\objectname{J2016+3657} &   74.76  & +0.98  & 0.44 & \phn\phd\phn0 & 1.02 & 0.63 & \phn9 & U0,WR,OB,RSN,Std,Pst &   \\
\objectname{J2020+4017} &   78.05  & +2.08  & 0.06 & \phn\phd\phn0 & 0.24 & \nodata & \phn9 & U0,OB,G,SNR,RSN,Std,Pst,Gem & $\gamma$ Cyg \\
\objectname{J2020$-$1545} &   28.09  & $-$26.62  & 0 & \phn\phd\phn0 & 1.30 & \nodata & \phn4 & U15,Std &   \\
\objectname{J2021+3716} &   75.58  & +0.33  & 0.36 & 0.03 & 0.83 & 0.71 & \phn9 & U0,WR,OB,Std,Pst,PWN &   \\
\objectname{J2022+4317} &   80.63  & +3.62  & 0.16 & \phn\phd\phn0 & 0.80 & 0.12 & \phn9 & U0,WR,OB,G,Std &   \\
\objectname{J2025$-$0744} &   36.90  & $-$24.38  & 0.87 & 0.42 & 1.81 & 3.14 & \phn4 & QSO & 2022$-$077 \\
\objectname{J2027+3429} &   74.08  & $-$2.36  & 0 & \phn\phd\phn0 & 0.43 & \nodata & \phn9 & U0,OB,Std,Pst &   \\
\objectname{J2033+4118} &   80.27  & +0.73  & 0.21 & \phn\phd\phn0 & 0.52 & 0.40 & \phn9 & U0,Of,OB,G,Std,Pst &   \\
\objectname{J2034$-$3110} &   12.25  & $-$34.64  & 1.18 & 0.55 & 2.05 & 1.43 & \phn5 & U30 &   \\
\objectname{J2035+4441} &   83.17  & +2.50  & 0.05 & \phn\phd\phn0 & 0.66 & \nodata & \phn9 & U0,G,Std,Pst &   \\
\objectname{J2036+1132} &   56.12  & $-$17.18  & 0 & \phn\phd\phn0 & 1.30 & \nodata & \phn5 & BLL & 2032+107 \\
\objectname{J2046+0933} &   55.75  & $-$20.23  & 0 & \phn\phd\phn0 & 0.80 & \nodata & \phn6 & U15,Std &   \\
\objectname{J2055$-$4716} &  352.56  & $-$40.20  & 0.93 & 0.34 & 2.09 & 1.59 & \phn3 & QSO & 2052$-$474 \\
\objectname{J2100+6012} &   97.76  & +9.16  & 0.13 & \phn\phd\phn0 & 0.70 & 0.03 & \phn6 & a & 2105+598 \\
\objectname{J2158$-$3023} &   17.73  & $-$52.25  & 0.59 & \phn\phd\phn0 & 1.66 & 0.75 & \phn3 & BLL & 2155$-$304 \\
\objectname{J2202+4217} &   92.59  & $-$10.44  & 0.80 & 0.02 & 1.57 & 0.85 & \phn8 & BLL & BL Lac \\
\objectname{J2206+6602} &  107.23  & +8.34  & 0 & \phn\phd\phn0 & 0.49 & \nodata & \phn5 & a & 2206+650 \\
\objectname{J2209+2401} &   81.83  & $-$25.65  & 0.91 & \phn\phd\phn0 & 1.86 & 0.36 & \phn5 & QSO & 2209+236 \\
\objectname{J2219$-$7941} &  310.64  & $-$35.06  & 0 & \phn\phd\phn0 & 0.89 & \nodata & \phn4 & U30,Std,Pst &   \\
\objectname{J2227+6122} &  106.53  & +3.18  & 0.20 & \phn\phd\phn0 & 0.82 & 0.21 & \phn5 & U0,OB,G,Std,Pst,PWN &   \\
\objectname{J2232+1147} &   77.44  & $-$38.58  & 0.48 & 0.18 & 0.97 & 1.22 & \phn7 & QSO & 2230+114 \\
\objectname{J2241$-$6736} &  319.81  & $-$45.02  & 0 & \phn\phd\phn0 & 1.54 & \nodata & \phn3 & U30 &   \\
\objectname{J2243+1509} &   82.69  & $-$37.49  & 1.06 & \phn\phd\phn0 & 1.79 & 0.86 & \phn7 & U30 &   \\
\objectname{J2248+1745} &   86.00  & $-$36.17  & 0.65 & \phn\phd\phn0 & 1.48 & 0.54 & \phn7 & U30,Std,Pst &   \\
\objectname{J2251$-$1341} &   52.48  & $-$58.91  & 1.30 & 0.73 & 2.19 & 4.04 & \phn5 & U30 &   \\
\objectname{J2254+1601} &   86.11  & $-$38.18  & 0.52 & 0.31 & 0.91 & 3.93 & \phn7 & QSO & 2251+158 \\
\objectname{J2255+1943} &   89.03  & $-$35.43  & 1.18 & 0.25 & 1.95 & 0.95 & \phn7 & a & 2250+1926 \\
\objectname{J2255$-$5012} &  338.75  & $-$58.12  & 0.50 & \phn\phd\phn0 & 2.00 & 0.33 & \phn2 & U30 &   \\
\objectname{J2314+4426} &  105.32  & $-$15.10  & 1.11 & 0.20 & 2.39 & 0.93 & \phn3 & U15,Std &   \\
\objectname{J2321$-$0328} &   76.82  & $-$58.07  & 1.36 & 0.78 & 2.27 & 2.46 & \phn5 & QSO & 2320$-$035 \\
\objectname{J2352+3752} &  110.26  & $-$23.54  & 1.37 & 0.72 & 2.43 & 2.23 & \phn4 & a & 2346+385 \\
\objectname{J2358+4604} &  113.39  & $-$15.82  & 0 & \phn\phd\phn0 & 0.61 & \nodata & \phn3 & QSO & 2351+456 \\
\objectname{J2359+2041} &  107.01  & $-$40.58  & 0.74 & \phn\phd\phn0 & 1.57 & 0.53 & \phn7 & QSO & 2356+196 \\
\enddata
\tablenotetext{a}{See text for comments about the variability
of 3EG J1746$-$2851.}
\tablecomments{The sources marked ``a'' in the Classes column
are low-confidence AGN identifications in the 3EG catalog.
For the other Classes, see Table~2.}
\end{deluxetable}

\end{document}